\documentclass[aps,pra,twocolumn,showpacs,preprintnumbers,amsmath,amssymb,superscriptaddress]{revtex4}

\usepackage{amsmath,amssymb}
\usepackage{bm}
\usepackage[latin1]{inputenc}
\usepackage{dcolumn}
\usepackage{graphicx}
\newcommand{\matrise}[1]{\begin{bmatrix} #1 \end{bmatrix}}

\begin{document}

\title{Security of quantum key distribution with bit and basis dependent detector flaws}

\author{Lars Lydersen}
\email{lars.lydersen@iet.ntnu.no}
\affiliation{Department of Electronics and Telecommunications, Norwegian University of Science and Technology, NO-7491 Trondheim, Norway}
\affiliation{University Graduate Center, NO-2027 Kjeller, Norway}

\author{Johannes Skaar}
\affiliation{Department of Electronics and Telecommunications, Norwegian University of Science and Technology, NO-7491 Trondheim, Norway}
\affiliation{University Graduate Center, NO-2027 Kjeller, Norway}

\date{\today}

\begin{abstract}
We consider the security of the Bennett-Brassard 1984 (BB84) protocol for Quantum Key Distribution (QKD), in the presence of bit and basis dependent detector flaws. We suggest a powerful attack that can be used in systems with detector efficiency mismatch, even if the detector assignments are chosen randomly by Bob. A security proof is provided, valid for any basis dependent, possibly lossy, linear optical imperfections in the channel/receiver/detectors. The proof does not assume the so-called squashing detector model.
\end{abstract}

\maketitle

\section{Introduction}
Quantum mechanics makes it possible to exchange a random bit string at a distance \cite{bennett1984,ekert1991,bennett1992,gisin2002}. In theory, the key distribution is secure, even if an eavesdropper Eve can do anything allowed by the currently known laws of nature \cite{mayers1996,mayers2001,lo1999,shor2000}.

In practical QKD systems there will always be imperfections. The security of QKD systems with a large variety of imperfections has been proved \cite{mayers1996,inamori2001,koashi2003,gottesman2004}. However, a QKD system is relatively complex, and loopholes and imperfections exist that are not covered by existing security proofs. A security loophole can be dealt with in two different ways: Either you modify the implementation, or you increase the amount of privacy amplification~\cite{csiszar1978} required to remove Eve's information about the key. The first approach, to modify the implementation, may often be done without decreasing the rate of which secret key can be generated. It may however increase the complexity of the implementation, which in turn may lead to new loopholes. The advantages of the second approach, to increase the amount of privacy amplification, are that the apparatus can be kept as simple as possible, and that existing implementations can be made secure with a software update. A drawback is clearly the reduced key rate, which is considered as a critical parameter in commercial QKD systems.

One of the imperfections to be considered in this paper, is called detector efficiency mismatch (DEM) \cite{makarov2006}. If an apparatus has DEM, Eve can control the efficiencies of Bob's detectors by choosing a parameter $t$ in some external domain. Examples of such domains can be the timing, polarization, or frequency of the photons \cite{makarov2006,fung2009}.

\begin{figure}
\includegraphics[width=6.5cm]{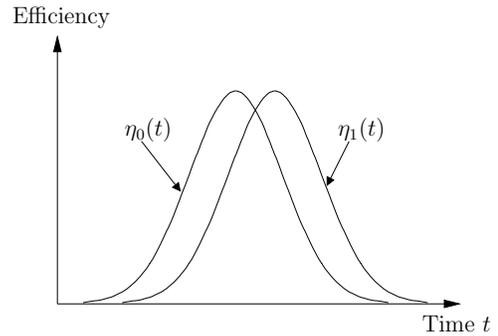}
\caption{\label{fig:efficiency_curves} An example of mismatched efficiency curves for two detectors in the time-domain. The functions $\eta_0(t)$ and $\eta_1(t)$ are the efficiencies of detector 0 and 1, respectively. The parameter $t$ can be used to parametrize other domains as well.}
\end{figure}

To be more concrete, consider DEM in the time-domain. In most QKD systems Bob's apparatus contains two single photon detectors to detect the incoming photons, one for each bit value. (Equivalently, two different detection windows of a single detector can be used for the two bit values (time-multiplexed detector).) Normally the detectors are gated in the time-domain to avoid high dark-counts. This means that electronic circuits are used to turn the detectors on and off, creating detection windows. Different optical path lengths, inaccuracies in the electronics, and finite precision in detector manufacturing may cause the detection windows of the two detectors to be slightly shifted, as seen in Fig.~\ref{fig:efficiency_curves}. The shift means that there exist times where the two detectors have different efficiencies.

Systems with DEM can be attacked with a faked-states attack \cite{makarov2006}. The faked-states attack is an intercept-resend attack where Eve does not try to reconstruct the original state sent by Alice, but rather exploit the imperfections in Bob's apparatus to hide errors. The faked-states attack can be adapted to the Scarani-Acin-Ribordy-Gisin 2004 (SARG04), Ekert, and Differential Phase Shift Keying (DPSK) protocols, in addition to BB84 \cite{makarov2008}. Another attack on systems with DEM is the time-shift attack \cite{qi2007}. In this attack Eve just selects the timing of each qubit randomly, thereby gaining information about the bit value when Bob announces which qubits were received and which were lost. The major advantage of the time-shift attack is that it does not introduce any quantum bit error rate (QBER). It has been demonstrated experimentally that the security of a commercially available QKD system can be compromised with a time-shift attack \cite{zhao2008}.

A frequently mentioned countermeasure for systems with DEM is called \emph{four-state Bob} \cite{nielsen2001,lagasse2005,makarov2006,qi2007}. In a phase-encoded QKD system, Bob chooses from four different phase settings $ \left\{ 0 , \pi/2, \pi, 3\pi /2 \right\} $ instead of only two $ \left\{ 0 , \pi/2 \right\} $. This will randomly assign the bit values 0 and 1 to the detectors (or the detection windows, in the case of one time-multiplexed detector) for each received state. Therefore Eve does not know which detector characteristics that corresponds to the 0 and 1 detectors.

However, as mentioned previously \cite{makarov2006,qi2007} Eve may use a large laser pulse attack \cite{ribordy2000,bethune2000,vakhitov2001,gisin2006} to read Bob's phase modulator settings. In a large pulse attack Eve uses a strong laser pulse to measure the reflections from either Alice's or Bob's apparatus. The setting of the phase modulator may give a signature on the reflections, enabling Eve to obtain the phase.

First assume that Eve is able to read Alice's modulator settings. Then Eve could obtain bit and/or basis information before the pulse enters Bob's apparatus, and therefore the security would be seriously compromised. Fortunately, Alice's implementation can easily be modified to avoid the large pulse attack. A setup with a coherent laser source contains an attenuator, and moving this to the end of the apparatus, as well as introducing an optical isolator, will put impossible requirements on Eve's laser \cite{vakhitov2001}. In ``plug-and-play'' systems Alice already uses a detector to monitor the input of her setup. Therefore a large pulse attack can easily be revealed by monitoring the intensity of the input.

In a straightforward implementation of BB84, the phase modulator setting in Bob's setup only contains basis information. It usually poses no security threat if Eve reads the basis, as she will get it during the public discussion anyway. One only has to avoid that Eve receives the basis information before the pulse enters Bob's apparatus. This can be taken care of by placing a properly long coil of optical fiber at the entrance of Bob's setup.

However, if the DEM loophole is patched with four-state Bob, the large pulse attack is dangerous, because it may give Eve information about the detector assignments. Modifying Bob's setup to avoid large pulse attacks is not an easy task. The most practical solution seems to be a beam splitter or an optical circulator combined with an intensity detector \cite{vakhitov2001}. Note that the key rate will suffer; the the input of Bob's setup is precious single photons. Also the setup gets more complex, which should be avoided as far as possible, to limit the number of ``hidden surprises''. It is therefore not obvious whether such modifications should be implemented, or whether the security should be regained with extra privacy amplification. Even though some systems implement four-state Bob, several of them lack countermeasures for a strong pulse attack on Bob's side. Therefore we will pursue the latter solution, i.e., we assume that Eve is able to read Bob's phase modulator setting after Bob's detection.

Security bounds state a unconditionally secure key rate, positive a range in some parameter(s). Ideally one should be able to prove the converse, namely that with the parameter(s) outside this range the QKD-system is provable insecure. Unfortunately this is not always simple. Usually there is a third range of the parameter(s) where it is not known whether the QKD-protocol is secure. For instance with perfect devices and one-way classical communication, the QKD-system is unconditionally secure for QBER $<$ 11 \% \cite{shor2000}, and provable insecure for QBER $>$ 14.6 \% \cite{fuchs1997}. Until the gap is closed the security bounds represent a lower bound on the secure key rate, and the best known attacks represent an upper bound.

Fung et al. found a security bound for QKD systems with DEM \cite{fung2009}. QKD systems with four-state Bob is proved to be secure, provided Eve cannot read Bob's phase settings with a large pulse attack. The security proof assumes the so-called squashing model \cite{gottesman2004}.

In this paper we first establish an upper bound for the secure key rate of QKD-system with DEM by presenting two powerful attacks, one of which even applies to implementations with four-state Bob (Section II). Then we will establish a lower bound for the secure key rate by providing a simple security proof of QKD systems with general, basis and bit dependent detector flaws (Section III), generalizing the proof by Fung et al. More precisely, any basis dependent, possibly lossy, linear optical imperfections in the channel and receiver are covered by the proof. For example, the proof covers mixing between all available optical modes, misalignments, mode-dependent losses, DEM, and any basis dependence of those effects. The proof is formulated for a decoy-state BB84 protocol and does not assume a squashing model. Finally, in Section IV we will examine some examples, including DEM, DEM with mode mixing, and DEM with misalignment.

\section{Security analysis: upper bound\label{fourstatebob}}

In this section we analyse two powerful attacks on systems with DEM. Such attacks are important because they establish a regime where QKD-systems with DEM is provable insecure. To analyze the attacks, for the moment we define $\eta = \max \left\{ \min_{t} \eta_1(t)/\eta_0(t), \min_t \eta_0(t)/\eta_1(t) \right\} \in [0,1]$, representing the smallest efficiency ratio available for both bit values. For individual attacks the secret key rate is given by \cite{csiszar1978,maurer1993} (given one-way classical communication)
\begin{equation}
   R = I(\alpha:\beta) - I(\alpha:\epsilon) \text{,}
   \label{eq:individual_rate}
\end{equation}
where $I(\cdot:\cdot)$ denotes mutual information and $\alpha$, $\beta$, and $\epsilon$ represent Alice's, Bob's and Eve's bits.

In the previous analysis of the faked-states attack \cite{makarov2006}, the attack was limited by the introduced QBER rather than Eve's insufficient knowledge about the key. By attacking only a fraction of the bits with the faked-states attack one can compromise the security for even higher values of $\eta$. The other fraction could be attacked with the time-shift attack \cite{qi2007} which introduces no QBER.

To tailor $E$, the QBER measured by Alice and Bob, the fraction $r$ attacked by the faked-states attack is given by
\begin{equation}
   r = \frac{E}{E_{\text{fs}}} = E\frac{1+3\eta}{2\eta} \text{,}
   \label{eq:fs_rate}
\end{equation}
where $E_{\text{fs}} = 2\eta/(1+3\eta)$ is the QBER introduced by the faked-states attack. The mutual information between Alice and Eve is given by
\begin{equation}
   \begin{split}
      I(\alpha:\epsilon) &= r I(\alpha:\epsilon)_{\text{fs}} + (1-r) I(\alpha:\epsilon)_{\text{ts}}\\
                         &= 1 - E - h(\frac{\eta}{1+\eta})\left(1 - \frac{1+3\eta}{2\eta}E \right) \text{,}
   \end{split}
   \label{eq:aemutual_comb}
\end{equation}
where $r$ is given in \eqref{eq:fs_rate} and $I(\alpha:\epsilon)_{\text{fs}} = 1 - E$ and $I(\alpha:\epsilon)_{\text{ts}} = 1 - h(\eta/(1+\eta))$ denote the mutual information in the faked-states and the time-shift attack, respectively, as given in Refs \cite{makarov2006,qi2007}. $h(\cdot)$ is the binary entropy function. Since Alice and Bob does not know how each bit is attacked, $I(\alpha:\beta)$ is simply given by $1 - h(E)$. The key rate \eqref{eq:individual_rate} thus becomes
\begin{equation}
   R = E + h(\frac{\eta}{1+\eta})\left( 1 - \frac{1 + 3\eta}{2\eta}E  \right) - h(E).
   \label{eq:key_rate_combined}
\end{equation}
Without considering DEM, Alice and Bob think that the key is secure when QBER $ < 11 \%$ (symmetric protocols with one-way classical communication \cite{shor2000}). Solving the equality $R = 0$, where $R$ is given by \eqref{eq:key_rate_combined}, and setting $E = 0.11$ gives $\eta = 0.215$.

The above combined attack is implementable with current technology. Up to $\eta = 0.160$ it represent an upper bound on the secure key rate (see Fig.~\ref{fig:eta_vs_qber}). However with four-state Bob, the attack is impossible since the faked-states attack requires knowledge of the bit-detector mapping before Bob receives the pulse.

For higher values of $\eta$ there exists an even more efficient attack. The optimal individual attack in the absence of imperfections is known \cite{fuchs1997}. Here Eve lets the qubit from Alice interact with a probe. After the basis is revealed, Eve's probe is in one of two non-orthogonal states \cite{fuchs1997}
\begin{subequations}
   \begin{align}
      |\xi_0 \rangle &= |0\rangle \\
      |\xi_1 \rangle &= \cos\varphi |0 \rangle + \sin\varphi |1\rangle \text{,}
   \end{align}
   \label{eq:optstates}
\end{subequations}
where $\varphi$ is related to the QBER by
\begin{equation}
   \cos\varphi = 1 - 2E \text{.}
   \label{eq:varphiE}
\end{equation}
Eve has to separate between $|\xi_0 \rangle$, corresponding to the bit value 0 at Alice, and $|\xi_1 \rangle$, corresponding to the bit value 1. The two states occur with an {\it a priori} probability 1/2.

In the presence of DEM, we improve the attack as follows: In addition to using a probe, Eve launches a time-shift attack. If Bob announces receipt, the probabilities of the two bit values is now $\left\{ 1/ \left( 1 + \eta \right) , \eta/ \left( 1 + \eta \right) \right\}$ according to the time-shift attack \cite{qi2007}. Then after the public discussion, Eve has to separate between the states \eqref{eq:optstates} with the {\it a priori} probabilities $\left\{ 1/ \left( 1 + \eta \right) , \eta/ \left( 1 + \eta \right) \right\}$. The optimal measurement is projective \cite{levitin1995}, and the probability $p$ of Eve measuring the correct bit value is found to be
\begin{equation}
   \begin{split}
      p &= \left( \frac{1}{1 + \eta} \right) \cos^2\left[ \frac{1}{2} \arctan\left(\frac{\sin 2\varphi}{ \frac{1}{\eta} - \cos 2\varphi}\right)\right] \\
        &+ \left( \frac{\eta}{1 + \eta} \right) \sin^2\left[\varphi + \frac{1}{2}\arctan\left(\frac{\sin 2\varphi}{ \frac{1}{\eta} - \cos 2\varphi}\right)\right] \text{,}
   \end{split}
   \label{eq:comb_suc_prob}
\end{equation}
where $\varphi$ is related to the QBER as in Eq. \eqref{eq:varphiE}.

Since Eve has probability $p$ to have the same bit value as Alice, $I(\alpha:\epsilon)$ is simply $1-h(p)$. $I(\alpha:\beta)$ is given by $1-h(E)$. The key rate \eqref{eq:individual_rate} for this improved optimal individual attack is thus
\begin{equation}
   \label{eq:key_rate_improved_fucs}
   R = h(p) - h(E) \text{,}
\end{equation}
where $p$ is given by \eqref{eq:comb_suc_prob}.

Without considering DEM, Alice and Bob think that the key is secure when QBER $ < 11 \%$. Solving the equality $R = 0$, where $R$ is given by \eqref{eq:key_rate_improved_fucs}, and setting $E = 0.11$ gives $\eta = 0.252$. In a commercial QKD system $\eta$ was found to be approximately $0.25$ (see Fig.~3 in \cite{zhao2008}) \footnote{Also note that the DEM found in this system is heavily asymmetric, and the attacks might be more powerful if optimized for asymmetric DEM.}. Therefore, this attack could be used to compromise the security of such QKD systems. Note that the attack does not require the bit-detector mapping until the post-processing step. Therefore systems patched with four-state Bob are vulnerable to the attack combined with a large pulse attack.

Note that the both attacks represent a substantial improvement compared to the previously published attacks which require $\eta < 0.066$ \cite{makarov2006}. Fig.~\ref{fig:eta_vs_qber} shows the range of $E,\eta$ which compromises security, and compares the two attacks.

\section{Security analysis: lower bound}
In this section we will prove the security of the BB84 protocol in the presence of bit and basis dependent detector flaws, and establish the secure key generation rate. We will prove the security in a general setting, lifting the so-called squashing model assumption. That is, Eve may send any multimode, photonic state, and Bob uses practical threshold detectors. Alice may use a single-photon source or phase-randomized faint laser pulses; in the latter case, Alice may use decoy states \cite{hwang2003,wang2005a,lo2005} to estimate photon-number dependent parameters. Alice's source is otherwise assumed perfect: It emits an incoherent mixture of photonic number states, randomly in logical modes ``0'' or ``1'', randomly in the $X$ or $Z$ bases, with no correlation between the bits, bases, and photon number statistics \cite{koashi2006}.

The state space accessible to Eve consists of the Fock space associated with all photonic modes supported by the channel. The channel and receiver is modeled as a basis-dependent quantum operation, $\mathcal C_Z$ and $\mathcal C_X$, in front of two threshold detectors. Here $Z$ and $X$ denote the bases chosen by Bob. Since reduced detector efficiencies can be absorbed into the quantum operations, we can let Bob's threshold detectors have perfect efficiency. Dark counts are attributed to Eve, and for double click events, Bob assigns a random value to his bit \cite{inamori2001,gottesman2004}.

In our security proof, the key condition of $\mathcal C_Z$ and $\mathcal C_X$ is that they are passive, in the sense of
\begin{equation}\label{vacuumcond}
|0\rangle \to |0\rangle,
\end{equation}
where $|0\rangle$ denotes the vacuum state of all modes. In other words, vacuum incident to all modes gives vacuum out. This condition is rather general; it includes all linear and nonlinear optical transformations of the modes supported by the channel.

For simplicity, however, we will restrict ourselves to linear optical imperfections. Bob's two detectors may still have different efficiencies, depending on the time, frequency, and/or polarization of the incoming states. Moreover, there may be imperfections in the channel and Bob's receiver. This can be described as arbitrary, square matrices $C_Z$ and $C_X$, acting on the \emph{channel modes} after Eve's intervention. The linear-optical property of $C_Z$ and $C_X$ is ensured from the fact that they are classical transformations (or transfer matrices) operating on the physical, photonic modes (e.g. temporal modes and polarization modes) rather than the total Fock space of the modes. Each mode can contain any photonic state such as number states or coherent states. Although $C_Z$ and $C_X$ have finite dimension, the associated, induced quantum operations $\mathcal C_Z$ and $\mathcal C_X$ operate on an infinite dimensional Fock space. We use the convention that Bob's basis selector is included in $C_X$ (see Subsection \ref{DEMexample}).

With singular value decomposition, we can write
\begin{equation}\label{decomp}
C_Z=U_ZF_ZV_Z C,
\end{equation}
where $U_Z$ and $V_Z$ are unitary operators, and $F_Z$ is a diagonal, positive matrix. In addition to the usual singular value decomposition, we have included an extra matrix factor $C$, governing losses and imperfections in the channel and/or receiver, independent of the basis chosen by Bob. The matrix $C$ may for example describe loss of the channel and time-dependent detector efficiencies common for the two detectors. The operator $C$ can be absorbed into Eve's attack, thus it never appears in the following analysis. The unitary operators $U_Z$ and $V_Z$ mix the modes together. For example, $V_Z$ is the result of sending the modes through a network isomorphic to the type in \cite{reck1994}. The diagonal matrix $F_Z$ represents the different efficiencies of the two detectors (in addition to basis-dependent absorptions in the receiver), and satisfies
\begin{equation}\label{FZdef}
|F_Z|^2=\text{diag}\matrise{\eta_{Z0}(t_1)&\eta_{Z1}(t_1)&\eta_{Z0}(t_2)&\eta_{Z1}(t_2)\ldots}.
\end{equation}
The parameters $t_j, j=1,2,\ldots$ label different modes. For example, $t_j$ may correspond to different temporal modes. In the absence of $U_Z$ and $V_Z$, $\eta_{Z0}(t_j)$ and $\eta_{Z1}(t_j)$ can be viewed as the efficiencies of detector 0 and 1 in the $Z$-basis. Otherwise the efficiencies $\eta_{Z0}(t_j)$ and $\eta_{Z1}(t_j)$ do not necessarily correspond to the detectors 0 and 1, respectively, nor to detection time $t_j$. However, the notation is selected as in the special case for intuition. Note that $F_Z$ may be represented as a collection of beam splitters with transmittivities $\eta_{Z0}(t_1)$, $\eta_{Z1}(t_1)$, and so forth. Then each mode is incident to its own beam splitter, and the vacuum state is sent into the other input.

\begin{figure}
\includegraphics[width=8.6cm]{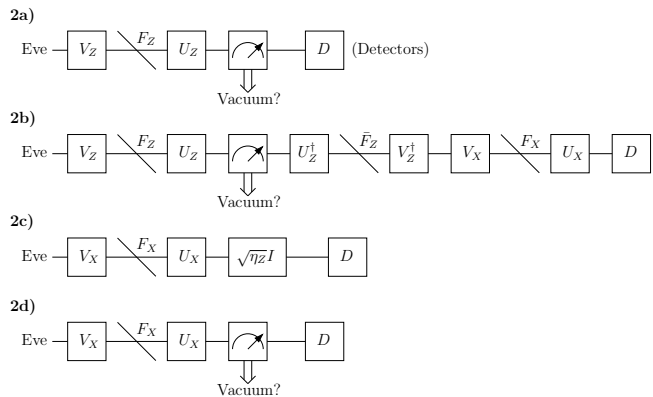}
\caption{\label{fig:models} a) Actual protocol. b) Estimation of Alice's virtual $X$-basis measurement. c) Simplification of Fig.~\ref{fig:models}b from Bob's point of view. d) Actual parameter estimation in the $X$-basis.}
\end{figure}

The resulting model is shown in Fig.~\ref{fig:models}a. In the model we have included an extra measurement, giving information to Eve whether the total state is equal to the vacuum $|0\rangle$. While this information actually comes from Bob, it is convenient to let Eve obtain this information from a separate measurement. Note that this extra vacuum measurement does not disturb Bob's measurement statistics for any basis choice.

We will prove security using Koashi's argument \cite{koashi2005,koashi2006a,koashi2006} which we briefly summarize here. In the BB84-like actual protocol Alice generates a large number of bipartite states, where her part consists of a qubit which she measures randomly in the $X$- or $Z$-basis. The other part of the pairs is sent to Bob via Eve. Bob measures what he receives from Eve randomly in two different bases, which we will refer to as the ``$X$-basis'' or the ``$Z$-basis''. For example, for polarization encoding Bob's two measurements should ideally correspond to threshold detectors in horizontal/vertical or $\pm 45^\circ$ polarization bases, with double clicks as random assignment. Alice and Bob discard all events where they used incompatible basis. Further he publicly announces receipt if he receives something different from vacuum. Let $Q_X$ and $Q_Z$ be the fractions of non-vacuum results in each basis. Alice and Bob compare their $X$-basis measurement results to estimate $Q_X$ and the error rate $E_X$. The $N$ states measured in the $Z$-basis yield $NQ_Z$ non-vacuum results. For these $NQ_Z$ events Alice's measurement result is the raw key.

The required amount of privacy amplification can be found as follows: imagine a virtual experiment where Alice measures the qubits for the raw key in the $X$-basis instead of the $Z$-basis. Bob tries to predict the result of Alice's virtual $X$-basis measurement. Bob does not perform such a prediction in practice; thus in this prediction we may let Bob do everything permitted by quantum mechanics, as long as he does not alter the information given to Eve. Let $H_{\text{virt}X}(A|B=\mu)$ denote the entropy of Alice's virtual $X$-basis measurement result, given measurement result $\mu$ in Bob's prediction. It turns out that $H_{\text{virt}X}(A|B=\mu)$ can be bounded using $E_X$ and $Q_X$, so assume that $H_{\text{virt}X}(A|B=\mu) \leq H$. Since the uncertainty about Alice $X$-measurement is less than $H$, the entropic uncertainty relation \cite{maassen1988} suggests that any prediction (including Eves prediction) of the measurement result of Alice $Z$-basis measurement will have at least $NQ_Z-H$ entropy. Thus Alice can extract $NQ_Z - H$ bits of secret key. Rigorously, this rate is found by concertizing the privacy amplification procedure by universal hashing. Although Koashi's original proof is formulated with an obsolete security definition based on accessible information, the proof can easily be adapted to a composable security definition \cite{renner2005a,ben-or2005,lo2007}.

Bob must ensure that he has an identical raw key. Since it does not matter to Eve what Bob does (as long as he gives Eve the same information), he measures the bits for the raw key in the $Z$-basis. Alice and Bob compares a subset of the raw key to find the error rate $E_Z$ (consuming some of the raw key, but negliable in the asymptotic limit), and Alice sends Bob $NQ_Zh(E_Z)$ bits of error correcting information consuming $NQ_Zh(E_Z)$ bits of previously established secret key. In the asymptotic limit $N \to \infty$ the net secure key generation rate becomes
\begin{equation}
   R_Z \geq 1 - \frac{H}{NQ_Z} - h(E_Z).
   \label{eq:koashi_rate}
\end{equation}

Note that $H$ is needed to ensure that Alice's key is secret, and this only requires $X$-basis parameters to be estimated by Alice and Bob. Thus there is no need to invoke the classicalization argument \cite{lo1999} regarding statistics of measurements involved in the simultaneous estimation of $E_X$ and $E_Z$.

For his prediction, Bob will use the virtual measurement in Fig.~\ref{fig:models}b. Bob first applies the unitary operator $U_Z^\dagger$, followed by the filter $\bar F_Z$, and the unitary operator $V_Z^\dagger$. Then he applies the operator $C_X=U_XF_XV_X$. Finally he performs an $X$-basis measurement. Note that we retain Eve's vacuum measurement and all components preceding it, so Eve obtains the identical information as in Fig.~\ref{fig:models}a. The matrix $\bar F_Z$ is diagonal, and is given by
\begin{equation}
\bar F_ZF_Z=\sqrt{\eta_Z} I,
\end{equation}
where
\begin{equation}\label{etadefgeneral}
\eta_Z=\min_{ij}\{\eta_{Zi}(t_j)\}.
\end{equation}
Similarly to $F_Z$, the filter $\bar F_Z$ is implementable by beam splitters acting separately on each mode. The largest element of $|\bar F_Z|^2$ is $1$, while the smallest element is $\eta_Z/\max_{ij}\{\eta_{Zi}(t_j)\}$.

To analyze how well Bob performs in his prediction, we will now simplify the system in Fig.~\ref{fig:models}b to determine Bob's measurement statistics. To do this, we introduce an extra vacuum measurement right before Bob's detectors, assuming nobody records the outcome. Clearly, Bob's measurement statistics are not altered by the presence of this extra measurement. The filter $U_XF_XV_XV_Z^\dagger\bar F_ZU_Z^\dagger$ obeys \eqref{vacuumcond}, being a linear optical transformation. As a result, we show in the appendix that the output state, after the extra vacuum measurement, is independent of the presence of Eve's vacuum measurement (i.e., the first vacuum measurement, after $U_Z$ in Fig.~\ref{fig:models}b). Thus, to estimate Bob's measurement statistics, we can remove Eve's vacuum measurement. We end up with the simplified system shown in Fig.~\ref{fig:models}c. Note that the simplified system is identical to the system in Fig.~\ref{fig:models}d, the actual protocol when Bob has chosen the $X$-basis, except for one thing: There is an extra, mode-independent absorption $\eta_Z$ in the channel. This fact will be used for estimating the performance of Bob's prediction.

To prove the security also for the multiphotonic case, we use the parameters $q_X^{(1)}$ and $e_X^{(1)}$ assumed known from the decoy state protocol. $q_X^{(1)}$ is the fraction of Bob's $X$-basis non-vacuum events that originate from single photons at Alice. $e_X^{(1)}$ is the QBER for single photon events in the $X$-basis (only single photons generate secure key). Consider the prediction in Fig.~\ref{fig:models}b-c. Let $NQ_Z$ be the number of states in the raw key. In a worst case, the number of detection events that originate from single photons at Alice, will be only $\eta_Z q_X^{(1)}Q_X N$, due to the filter $\sqrt\eta_Z I$ (note that $\eta_Z Q_X < Q_Z$). For each of these events Bob's entropic uncertainty about Alice's bit is (asymptotically) $h(e_X^{(1)*})$, where $e_X^{(1)*}$ is the associated error rate. We note that $e_X^{(1)*}$ is not measured in the actual protocol; it will rather be estimated below. For the events lost in the filter $\sqrt\eta_Z I$, Bob's entropic uncertainty about Alice's bit is 1, since he has no detection result. Summarizing, Bob's entropic uncertainty about Alice's $Q_Z N$ bits (corresponding to the number of detection events in Fig.~\ref{fig:models}a) is at most $H=Q_Z N-\eta_Z q_X^{(1)}Q_X N[1-h(e_X^{(1)*})]$. In our analysis we have ignored the events associated with Alice sending the vacuum state \cite{koashi2006}; their contribution will only give a marginally larger rate. From \eqref{eq:koashi_rate} the secure key rate becomes
\begin{equation}
R_Z = -h(E_Z) + \eta_Z q_X^{(1)}Q_X/Q_Z\left[1-h(e_X^{(1)*})\right].
\end{equation}

It remains to bound the parameter $e_X^{(1)*}$, which is the QBER for single photon events in the estimation Fig.~\ref{fig:models}b-c. Recall that $e_X^{(1)}$ is the estimated QBER for single photon events in the $X$-basis, Fig.~\ref{fig:models}d. The only difference between the setup in Fig.~\ref{fig:models}c and Fig.~\ref{fig:models}d is the filter $\sqrt\eta_Z I$, which represent identical absorption in all modes. However, the removal of detection events by this filter is dependent on the photon number, so $e_X^{(1)*}\neq e_X^{(1)}$ in general \footnote{Note that although Alice send a single photon for a particular event, Eve may send any state.}. To bound $e_X^{(1)*}$ we use the fact that the filter only alter the detection statistics by removing detection events. (An exception occurs for the few coincidence counts; these can be taken into account easily.) In a worst case,
\begin{equation}\label{erroramplification}
e_X^{(1)*}\leq \frac{e_X^{(1)}}{\eta_Z(1-e_X^{(1)})+e_X^{(1)}}\leq e_X^{(1)}/\eta_Z.
\end{equation}
Putting these results together, we obtain the secure key generation rate
\begin{equation}\label{rate}
R_Z \geq -h(E_Z)+\eta_Z q_X^{(1)}Q_X/Q_Z \left[1-h(e_X^{(1)}/\eta_Z)\right].
\end{equation}
A similar result holds when Alice and Bob have chosen the $X$-basis in the actual protocol:
\begin{equation}\label{rateX}
R_X \geq -h(E_X)+\eta_X q_Z^{(1)}Q_Z/Q_X \left[1-h(e_Z^{(1)}/\eta_X)\right].
\end{equation}
Ineqs. \eqref{rate} and \eqref{rateX} are valid for any basis and bit dependence of the channel and receiver/detectors, as long as the imperfections ($C_Z$ and $C_X$) can be described as possibly lossy, linear optical operators acting on the photonic modes.

To compare our result \eqref{rate} to that of Ref. \cite{fung2009}, we let Alice only send single photons. The rate then becomes
\begin{equation}
R \geq -h(E)+\eta [1-h(E/\eta)], \label{ratefung}
\end{equation}
where we have assumed symmetry between the bases, and therefore omitted the $Z$ and $X$ subscripts. The rate \eqref{ratefung} coincides with the rate found in \cite{fung2009} (see Subsection \ref{fungexample} for a discussion on how to identify $\eta$). Note, however, that \eqref{ratefung} is a stronger result in the sense that it applies to any basis-dependent linear optical imperfections, not only the case where $U_{Z,X}=I$, and $V_{Z,X}$ do not mix modes associated with different logical bits. Also it does not require the squashing model assumption.

Under the assumption that Eve only sends single photons, it is easy to realize that \eqref{erroramplification} can be replaced by $e_X^{(1)*}= e_X^{(1)}$. Then \eqref{ratefung} is improved to
\begin{equation}
R \geq -h(E)+\eta [1-h(E)]. \label{ratefungsingleph}
\end{equation}

\begin{figure}
\includegraphics[width=8.6cm]{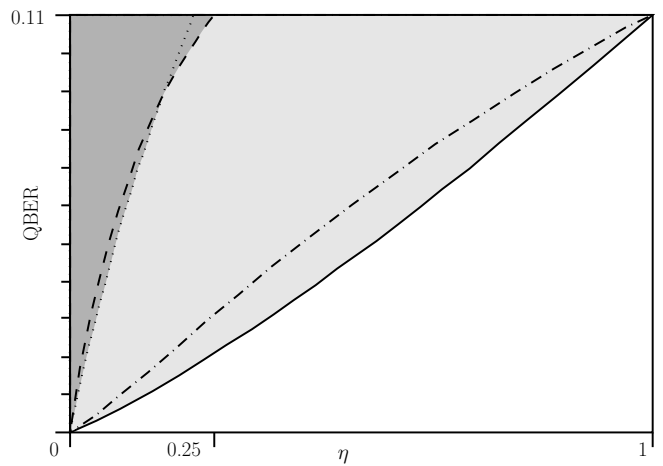}
\caption{\label{fig:eta_vs_qber} Security bounds when Alice sends single photons ($q_Z^{(1)}=q_X^{(1)}=1$), assuming symmetry between the bases. The bounds are found by setting the associated key generation rates equal to zero. Solid line: General security bound, as resulting from \eqref{ratefung}. Dash-dotted line: Security bound \eqref{ratefungsingleph} assuming Eve sends single photons. Dashed line: The improvement of the optimal individual attack from Section~\ref{fourstatebob}, as resulting from \eqref{eq:key_rate_improved_fucs}. Dotted line: The combined attack from Section~\ref{fourstatebob}, as resulting from~\eqref{eq:key_rate_combined}. For the attacks it is assumed that the DEM is equal for the two bit values. The dark grey region is proved to be insecure while the white region is proved to be secure with extra privacy amplification. The light grey region should be assumed insecure.}
\end{figure}

Fig.~\ref{fig:eta_vs_qber} shows the security bounds resulting from \eqref{ratefung} and \eqref{ratefungsingleph} when the right-hand side is set equal to zero.

\section{Examples}

\subsection{DEM in the time-domain}\label{DEMexample}
Consider the case where Bob's detectors have time-dependent efficiencies, as indicated in Fig.~\ref{fig:efficiency_curves}. We assume that the efficiencies are independent of the basis chosen by Bob ($F_X=F_Z$). The channel and receiver are otherwise assumed perfect, except for a background loss $C$. The background loss may be mode dependent, but independent of the basis chosen by Bob.

With these assumptions, we may take $C_Z=F_Z C$ and $C_X=F_XH C=F_ZH C$, where $H$ is a block-diagonal matrix consisting of $2\times 2$ Hadamard matrices $H^{(2)}$, interchanging the bases $Z$ and $X$ for each time:
\begin{equation}
H=\text{diag}\matrise{H^{(2)}&H^{(2)}&H^{(2)}&\ldots}.
\end{equation}
To maximize the secure key rate, as much as possible of the detector flaws should be absorbed into $C$. Therefore, we factorize
\begin{equation}\label{factorizeloss}
F_Z=FF',
\end{equation}
where
\begin{equation}\label{Fpdef}
F'^2=\text{diag}\matrise{\eta'(t_1)&\eta'(t_1)&\eta'(t_2)&\eta'(t_2)\ldots},
\end{equation}
and $\eta'(t_j)=\max\{\eta_{Z0}(t_j),\eta_{Z1}(t_j)\}$. Noting that $F'$ and $H$ commute, we can absorb $F'$ into $C$. The remaining diagonal matrix $F$ then has the role of $F_Z$ (and $F_X$) in the security proof. The parameter $\eta_Z=\eta_X$ to substitute into the secure key generation rate \eqref{rate} is therefore the minimum diagonal element of $|F|^2$:
\begin{equation}\label{etadef}
\eta_Z=\min_t\min\left\{\frac{\eta_{Z0}(t)}{\eta_{Z1}(t)},\frac{\eta_{Z1}(t)}{\eta_{Z0}(t)}\right\}.
\end{equation}

\subsection{DEM and restricted mode mixing}\label{fungexample}
Consider the case treated by Fung et al. \cite{fung2009}, where there is no mixing between modes associated with different logical bits. Then $C_Z$ can be written in block diagonal form
\begin{equation}
C_Z = \matrise{C_0 & 0 \\ 0 & C_1} C,
\end{equation}
provided we reorder the modes as in
\begin{equation}\label{FZdef2}
|F_Z|^2=\text{diag}\matrise{\eta_{Z0}(t_1)&\eta_{Z0}(t_2)&\ldots\ \eta_{Z1}(t_1)&\eta_{Z1}(t_2)&\ldots},
\end{equation}
to be compared to \eqref{FZdef}. As in Ref. \cite{fung2009} we assume basis independence in the sense
\begin{equation}
C_X = \matrise{C_0 & 0 \\ 0 & C_1} H C.
\end{equation}
Here,
\begin{equation}
H = \frac{1}{\sqrt 2}\matrise{I & I \\ I & -I},
\end{equation}
with the present choice of mode order. We assume that $C_Z$ is nonsingular. (Otherwise, the secure key generation rate would be zero.)

We should associate as much as possible of the imperfections to the common channel operator $C$. Let the singular-value decomposition of $C_0C_1^{-1}$ be $usv$, where $u$ and $v$ are unitary matrices, and $s$ is diagonal and positive. Let $\lambda^2$ be the maximum of $\max s$ and $\max s^{-1}$. Factorize
\begin{equation}
C_Z = \lambda\matrise{u s^{1/2} & 0 \\ 0 & v^\dagger s^{-1/2}}\frac{1}{\lambda}\matrise{s^{-1/2}u^\dagger C_0 & 0 \\ 0 & s^{1/2}vC_1} C.
\end{equation}
Defining
\begin{equation}
C' = \frac{1}{\lambda}\matrise{s^{-1/2}u^\dagger C_0 & 0 \\ 0 & s^{1/2}vC_1},
\end{equation}
and noting that $s^{-1/2}u^\dagger C_0=s^{1/2}vC_1$, we have $C'H=HC'$. This gives
\begin{subequations}\label{CZCX}
\begin{align}
C_Z &= \lambda\matrise{u s^{1/2} & 0 \\ 0 & v^\dagger s^{-1/2}} C'C, \\
C_X &= \lambda \matrise{u s^{1/2} & 0 \\ 0 & v^\dagger s^{-1/2}}H C'C.
\end{align}
\end{subequations}
Similarly to the reasoning in Section III, Bob applies a virtual filter to transform $C_Z$ into an operator proportional to $C_X$. Applying
\[  \frac{1}{\lambda}\matrise{u s^{1/2} & 0 \\ 0 & v^\dagger s^{-1/2}}H \frac{1}{\lambda}\matrise{s^{-1/2}u^\dagger  & 0 \\ 0 & s^{1/2}v}, \]
the operator $C_Z$ is transformed into $C_X/\lambda^2$. Following Section III, $\sqrt\eta=1/\lambda^2$. This gives
\begin{equation}\label{etasdef}
\sqrt\eta=\min(\min s,\min s^{-1}).
\end{equation}
Equivalently, $\eta$ is the minimum value of the eigenvalues and inverse eigenvalues of $C_0C_1^{-1} (C_0C_1^{-1})^\dagger=C_0 (C_1^\dagger C_1)^{-1}C_0^\dagger$. This $\eta$ should be substituted into \eqref{rate} to find the secure key generation rate.

The parameter $\eta$ can be measured as follows. For single photon input in a given superposition $\psi$ of logical ``0'' modes, the probability of a click in detector $0$ is given by $ \psi^\dagger C_0^\dagger C_0 \psi $. Similarly, we may use the identical superposition $\psi$ of ``1'' modes to find the detection probability of detector $1$. Note that $\psi$ denotes a classical field vector, where each element corresponds to a separate mode. The parameter $\eta$ turns out to be equal to the minimum detection probability ratio
\begin{equation}
   \eta = \min \left(\min_{\psi } \frac{ \psi^\dagger C_0^\dagger C_0 \psi }{ \psi^\dagger  C_1^\dagger C_1  \psi }, \min_{\psi } \frac{ \psi^\dagger C_1^\dagger C_1 \psi }{ \psi^\dagger  C_0^\dagger C_0  \psi } \right) \text{.}
\end{equation}
In other words, $\eta$ is given by the minimum efficiency mismatch ratio for all superpositions of input modes.

To see this, let $u s^2 u^\dagger$ be the spectral decomposition of $C_0 (C_1^\dagger C_1)^{-1}C_0^\dagger$. Then we have $C_0^{-1 \dagger} (C_1^\dagger C_1) C_0^{-1} = u s^{-2} u^\dagger$, and
\begin{equation}\label{psiratio}
\begin{split}
   \frac{ \psi^\dagger C_1^\dagger C_1 \psi }{ \psi^\dagger  C_0^\dagger C_0  \psi }
   &= \frac{ \psi'^\dagger C_0^{-1 \dagger} C_1^\dagger C_1 C_0^{-1} \psi' }{ \psi'^\dagger  \psi' } \\
   &= \frac{ \psi'^\dagger u^\dagger s^{-2} u \psi' }{ \psi'^\dagger \psi' } \\
   &=  s^{-2} \text{.}
\end{split}
\end{equation}
Combining \eqref{etasdef} and \eqref{psiratio} gives the desired result.

\subsection{DEM and misalignments}
In addition to the detector efficiency mismatch in Subsection \ref{DEMexample}, suppose that Bob's detectors are misaligned. The misalignments may be dependent on Bob's choice of basis, and are described by unitary matrices $V_Z$ and $V_X$. This gives the channel operators $C_Z=F_ZV_ZC$ and $C_X=F_XV_XHC$. Assuming no coupling between different temporal modes (no multiple reflections), $V_Z$ and $V_X$ are block-diagonal matrices. For example,
\begin{equation}
V_Z=\text{diag}\matrise{V_1^{(2)}&V_2^{(2)}&V_3^{(2)}&\ldots},
\end{equation}
where $V_j^{(2)}$ are unitary $2\times 2$ matrices. Here we have used the same order of modes as in the original definition \eqref{FZdef}. Taking $F_X=F_Z$ and factorizing as in Subsection \ref{DEMexample}, we find that the parameter $\eta_Z=\eta_X$ again is given by \eqref{etadef}. The secure key generation rate is then found from \eqref{rate}.

If there is coupling between modes associated with different $t$'s (in addition to the misalignment), we must retain the general definition of $\eta_Z$ in \eqref{etadefgeneral}. For unnormalized detection efficiencies, this definition can be rewritten
\begin{equation}\label{etadefgeneralunn}
\eta_Z=\frac{\min_{i,t}\{\eta_{Zi}(t)\}}{\max_{i,t}\{\eta_{Zi}(t)\}}.
\end{equation}
Eq. \eqref{etadefgeneralunn} is obtained by absorbing the maximum detector efficiency $\max_{i,t}\{\eta_{Zi}(t)\}$ into $C$. Omitting the requirement $F_X=F_Z$, \eqref{etadefgeneralunn} must be rewritten as
\begin{equation}\label{etadefgeneralunngen}
\eta_Z=\frac{\min_{i,t}\{\eta_{Zi}(t)\}}{\max\left(\max_{i,t}\{\eta_{Zi}(t)\},\max_{i,t}\{\eta_{Xi}(t)\}\right)}.
\end{equation}

\subsection{Characterizing DEM of Bob's receiver}
To estimate the secure key generation rate, Bob must characterize his receiver to find $\eta_Z$ and $\eta_X$ (or $\eta\equiv\min\{\eta_Z,\eta_X\}$). We note that rather different results are obtained dependent on whether or not there are coupling between different modes. For the case of DEM in the time-domain, since it is difficult to eliminate multiple reflections in Bob's receiver, a conservative approach is to use \eqref{etadefgeneralunngen}.

For the case with gated detectors, the efficiencies approach zero at the edges of the detection window. When there are coupling between different temporal modes, the resulting key generation rate will therefore be close to zero. Even if no such coupling is present, the key generation rate may approach zero, since at the edges of the detection window the efficiency ratio may be very small. (Although the average detection probability at the edges may be small, Eve may compensate this by replacing the channel by a more transparent one, or by increasing the power of her pulses \cite{makarov2006}.) A possible solution may be that Bob monitors his input signal at all times, to ensure that Eve does not send photons outside the central part of the window.
Then $\eta$ can be obtained by measuring the minimum and maximum detection efficiency for (superpositions of) modes with times inside this central part.

Such a measurement may be cumbersome due to many degrees of freedom of the possible inputs. Alternatively, one could specify the maximum possible amount of mode coupling in the system, and use this information to lower bound $\eta$. Suppose that the maximum (power) coupling from one mode $j$ to all other modes is $\delta$. Then the unitary matrix $V_Z$ satisfies $\sum_{i, i\neq j}|V_{ij}|^2 < \delta$ in addition to $\sum_{i}|V_{ij}|^2=1$, omitting the subscript $Z$ for clarity. Let $|f_j|^2$ be the $j$th diagonal element of $F_Z$. By measuring the detection efficiency when photons are incident to the $j$th mode, we obtain $\sum_{i}|V_{ij}|^2|f_i|^2 = |f_j|^2 + \sum_{i,i\neq j}|V_{ij}|^2\left(|f_i|^2-|f_j|^2\right)$. Hence, the elements $|f_j|^2$ can be found from the detection efficiency as a function of $j$ of the incident mode, up to an error $\left|\sum_{i,i\neq j}|V_{ij}|^2\left(|f_i|^2-|f_j|^2\right)\right|<\delta$. A lower bound of $\eta$ is therefore
\begin{equation}\label{etameasurement}
\eta>\frac{\min_{t,\text{basis,bit}}(\text{detection efficiency})-\delta}{\max_{t,\text{basis,bit}}(\text{detection efficiency})+\delta}.
\end{equation}
The required measurement is to obtain the detection efficiency as a function of $t$ and logical bit value for both bases. For detection efficiency mismatch in the time-domain the test pulses should be sufficiently short, in order to capture all details. An upper bound of the parameter $\delta$ may be estimated from the (worst case) multiple reflections and misalignment's that may happen in the system.

\section{Discussion and conclusion}

In this work we have proved the security of BB84 in the presence of any basis dependent, possibly lossy, linear optical imperfections in the channel and receiver/detectors. The security proof thus covers a combination of several imperfections: Detection efficiency mismatch, misalignments, mixing between the modes, multiple reflections, and any basis dependence of those effects. Contrary to most previous security proofs, this proof does not require a squashing detector model.

A specific implementation of a QKD system may have several different imperfections. Ideally there should be a universal security proof with a set of parameters that cover all (worst case) imperfections and tolerances of the equipment. We have made a step towards this goal by describing generic imperfections at the detector, and by providing a compact proof, which may hopefully prove useful for an even more general description.

We have established an upper bound for the secure key rate by providing two powerful attacks. One of the attacks may be applied to systems even with the four-state Bob patch, and this demonstrates the seriousness of the detection efficiency loophole. This attack is based on a combination of an optimal individual attack, a time shift attack, and a large pulse attack. As a consequence of such types of attacks, the key generation rate may not increase substantially as a result of the four-state Bob patch. A possible countermeasure is to use the general bounds \eqref{rate} and \eqref{rateX} for estimating the required amount of privacy amplification.

\begin{acknowledgments}
We would like to thank Vadim Makarov for comments and discussions. Financial support from NTNU and the Research Council of Norway (grant no.\ 180439/V30) is acknowledged.
\end{acknowledgments}

\appendix
\section{Properties of vacuum measurement}
\begin{figure}[ht]
\includegraphics[width=8.6cm]{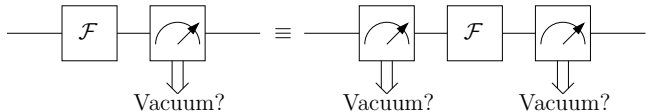}
\caption{\label{fig:vacuumtrick} The statistics and output state of the vacuum measurement after $\mathcal F$ is not changed by the introduction of a vacuum measurement before $\mathcal F$.}
\end{figure}

Let $\{|n\rangle\}$ be an orthonormal basis for a state space of interest. We refer to the state $|0\rangle$ as the ``vacuum state of all modes'', although it could in principle be any fixed, pure state. A vacuum measurement is a projective measurement with projectors $P=|0\rangle\langle 0|$ and $I-P$. We claim that if $\mathcal F$ is any quantum operation satisfying \eqref{vacuumcond}, i.e.,
\begin{equation}\label{vacuumcond2}
\mathcal F(|0\rangle\langle 0|)=|0\rangle\langle 0|,
\end{equation}
the presence of a vacuum measurement before $\mathcal F$ does not change the statistics and output state of a vacuum measurement after $\mathcal F$, see Fig. \ref{fig:vacuumtrick}.

This result can be proved by using the fact that any quantum operation can be viewed as a unitary transformation on an extended state space, with a standard state $|0\rangle_\text{aux}$ as auxiliary input. Due to \eqref{vacuumcond2}, we can assume that the unitary transformation transforms
\begin{equation}
\label{vacuumcond3}
|0\rangle\otimes|0\rangle_\text{aux}\to|0\rangle\otimes|0\rangle_\text{aux},
\end{equation}
with no loss of generality.

Consider the right-hand side of the identity (Fig. \ref{fig:vacuumtrick}). Let $P_\text{aux}=|0\rangle_\text{aux}\langle 0|_\text{aux}$. A vacuum measurement at the input can now be described as a projective measurement with $P\otimes P_\text{aux}$ and $I-P\otimes P_\text{aux}$, since the auxiliary input is fixed at $|0\rangle_\text{aux}$. Clearly, it does not matter if we measure the auxiliary output with projectors $P_\text{aux}$ and $I-P_\text{aux}$. In total, the extended measurement at the output is described by projectors $P\otimes P_\text{aux}$, $P\otimes (I-P_\text{aux})$, $(I-P)\otimes P_\text{aux}$, and $(I-P)\otimes (I-P_\text{aux})$. Transforming the projector $P\otimes P_\text{aux}$ backwards, we find that the corresponding projector at the input is $P\otimes P_\text{aux}$. In other words, the extended vacuum measurement at the output contains the vacuum measurement at the input, so the latter is redundant.

\bibliography{bibtex_library}

\end{document}